------------------------------------------------------------------------------
\documentstyle[12pt]{article}
\begin{document}
\title
{Long-Range Order of the Site-Random Spin Glass Model\footnote{
J. Phys. Soc. Jpn. {\bf 64} (1995) No.9}
}
\author{Yukiyasu O{\sc zeki} and Yoshihiko N{\sc onomura}}
\date{}
\maketitle
\begin{center}
Department of Physics, Tokyo Institute of Technology, \\
Oh-okayama 2-12-1, Meguro-ku, Tokyo 152, Japan
\end{center}

%
%
%
\def\Jij{J_{ij}}
\def\rJ{{\rm J}}
\def\cA{c}
\def\pF{p_{\rm FR}}
\def\ccF{c_{\rm f}}
\def\ccA{c_{\rm a}}
\def\aU{a_{\rm U}}
\def\aS{a_{\rm S}}
\def\aR{a_{\rm R}}
\def\WUL{W_{{\rm U},L}}
\def\WSL{W_{{\rm S},L}}
\def\WRL{W_{{\rm R},L}}
\noindent
\abstract
The site-random Ising spin glass model is investigated.
We find a rigorous symmetry for the SG correlation
and the free energy,
which provides some restrictions in the phase diagram.
Using the defect energies calculated by the numerical transfer matrix method,
we obtain evidence for the existence of the SG phase
in the two-dimensional Ising system.
We suggest
that the transitions from the FM and
the AF phases in the ground state in this model
can be explained by the percolated-cluster picture,
which is quite different from the frustration picture in the
conventional $\pm J$ model.\\

\endabstract
\bigskip
Since the pioneering work by Edwards and Anderson (EA),\cite{EdwarA75}
the spin glass (SG) problem has been investigated mainly
in terms of bond-random models,\cite{BindeY86}
which are generically called the EA models.
Up to the mean-field level, the theoretical picture of the EA model has been
established, and the system exhibits
a typical many-valley structure in the phase space.
Since the upper critical dimensions of these models are quite large,
such a mean-field picture might be modified in a physical situation.
Although the SG transition has been confirmed in
the 3d Ising SG models,
it has been denied in the 2d Ising \cite{BhattY85}
and the 3d Heisenberg models.\cite{OliveY86,MorriC86,OzekiN92,MatsuI91}
The real SG transitions are now
interpreted theoretically
as, for example,
a crossover to the Ising SG\cite{OliveY86,MorriC86,MatsuI91}
or induced order from the Ising chiral SG.\cite{Kawamu92}

Recently, Shirakura and Matsubara\cite{ShiraM93}
investigated the site-random Ising SG model numerically for $d=2$ and $d=3$,
and found evidence
for the existence of the SG phase even for $d=2$.\cite{ShiraM94}
This behavior is quite different from that of the EA model.
Since most of the SG materials can be described very well
using site-random models,
this model is more realistic than the EA-type models
such as the $\pm J$ Ising model.
In particular, an SG material, $\rm Fe_{\it x} Mn_{\it 1-x} TiO_3$
is closely related to it.\cite{KatorI94}
Thus, this model is expected to clarify the problems and
the difficulties in
the conventional SG theory based on the bond randomness.
In the present article,
we show characteristic behavior of the site-random model
not only numerically but also analytically.
Although this model has no gauge symmetry in contrast to the $\pm J$
model,\cite{Nishim81,Kitata92,OzekiN93,Ozeki95}
we find another type of symmetry which provides a restriction of
the structure of the phase diagram.
We also find numerically typical orderings of the ferromagnetic (FM) and
the antiferromagnetic (AF) phases in $d=2$.
These results reveal a physical interpretation of the SG as well as
the FM (AF) orderings.

The present site-random model consists of
two kinds of magnetic ions (A and B) randomly
distributed on all lattice sites
with a fixed concentration, $\cA$, of the A-ion.
Let us consider an Ising spin system
${\cal H} = -\displaystyle\sum_{\langle ij\rangle} J_{ij} S_i S_j$
with nearest-neighbor interactions
on a $d$-dimensional hypercubic lattice.
The exchange interaction between $i$-th and $j$-th sites, $\Jij$,
is determined by the species of the ions on these sites;
$\Jij =-J$ if both are B-ions, and $J_{ij}=+J$ otherwise.
Mathematically,
we associate an independent quenched random variable
$\omega_i$ with each site according to the species of ions;
$\omega_i =1$ and $\omega_i =-1$
represent the A-ion and the B-ion, respectively.
Then, the exchange interaction is expressed as
$J_{ij}={J\over 2}(1+\omega_i +\omega_j -\omega_i \omega_j )$.
The average of the ion configuration is denoted as
$[\cdots ]_\cA = \displaystyle\sum_{\{\omega_i =\pm 1\}}
\cdots \prod_i P_\cA (\omega_i )$,
where the distribution function of $\omega_i$ is given by
$P_\cA (\omega_i ) = \cA\delta (\omega_i -1) +(1-\cA )\delta (\omega_i +1)$.

When the concentration $\cA$ is close to unity (zero),
the FM (AF) phase appears in the low-temperature region
for $d\ge 2$.
At intermediate concentrations, $\cA\sim 1/2$,
the SG phase is expected to appear for sufficiently high dimensions.

The probability $\pF$ that one particular plaquette is frustrated
is given by
$\pF =4\cA^2 (1-\cA )^2 $.
Although the positions of frustrations are correlated with each other,
we numerically find that this is a good approximation for
the concentration of frustrations, at least in two dimensions.
Similarly,
the probability $p$ that one particular bond takes $\Jij =+J$
is given by
$p=2\cA -\cA^2$.
We also find that this is a good approximation of the
concentration of $+J$  bonds.
Note that $p$ is greater than $1/2$ at $\cA =1/2$,
and becomes $1/2$ when $\cA =1-1/\sqrt{2} \approx 0.293$.
This is related to the absence of explicit symmetry
with respect to $c=1/2$.

In the $\pm J$ Ising model, the phase diagram is symmetric
with respect to $p=1/2$ when the FM and AF phases are identified.
This is due to the invariance of the Hamiltonian
in terms of the successive transformations
$S_i \to (-1)^i S_i$ and $\Jij \to -\Jij$ for all sites and bonds.
Here $(-1)^i$ gives a negative sign
when the $i$-th site is located in one of the sublattices.
This invariance means
that the FM correlation function at $p$
is identical with the AF correlation function at $1-p$,
and the SG correlation function
is invariant under the transformation $p \leftrightarrow 1-p$.

In the present model, the Hamiltonian is invariant under the
successive transformations
$S_i \to (-1)^i S_i$,
$S_i \to \omega_i S_i$ and
$\omega_i \to -\omega_i$
for all sites
(the order of operations is arbitrary).
The first one represents the sublattice flip,
the second one flips all spins on B-ions,
and the last one switches the labels of the A-ion and the B-ion.
Note that the last one changes the ion distribution
$P_\cA (\omega )$ to $P_{1-\cA} (\omega)$.
Then, the free energy as well as
the SG correlation function,
$\left[\langle S_i S_j\rangle^2 \right]_\cA$,
is invariant under
the transformation, $c \leftrightarrow 1-\cA$,
while the FM and the AF correlation functions,
$\big[\langle S_i S_j \rangle\big]_\cA$ and
$(-1)^{i-j}\big[\langle S_i S_j \rangle \big]_{1-\cA}$,
have no such symmetry.
In fact, they satisfy
$\left[\langle S_i S_j \rangle \right]_\cA =
(-1)^{i-j}
\left[\langle S_i S_j \rangle \omega_i \omega_j \right]_{1-\cA}$.
Since the SG order parameter does not vanish
even in the FM and AF phases,
the above symmetry rigorously shows that
the SG phase must exist in the excess region when
the FM and AF phases are not symmetric (see Fig. 1),
and that the boundary of the paramagnetic phase,
the onset of the SG order parameter, is symmetric
with respect to $\cA =1/2$.

When the bond concentration is close to zero or unity,
one can approximate the present system
by the $\pm J$ model with
the bond concentration
$p=2\cA -\cA^2$ (the $\pm J$ mapping).
The FM (AF) critical point has been obtained numerically
as $p\approx 0.89$ ($p\approx 0.11$)
in the ground state in two dimensions.\cite{OzekiN87,Ozeki90,UenoO91}
Using the above mapping,
the corresponding critical concentrations of the present model
are estimated as
$\ccF \approx 0.668$ for the FM critical point, and
$\ccA \approx 0.057$ for the AF critical point.
Of course, this approximation is not good around the critical concentration,
where we should take the correlation of bonds into account.

In order to estimate these critical concentrations precisely,
we calculate the defect energy in the ground state in $d=2$
using the numerical transfer matrix method introduced
previously.\cite{Ozeki90,UenoO91}
Three kinds of boundary conditions (BCs) are applied;
the (usual) uniform BC conjugate to the FM order,
the staggered BC conjugate to the AF order, and
the replica BC\cite{Ozeki93b} conjugate to all kinds of long-range
order.\footnote{
The fixed spins on the boundary are modified from
those in ref. 19, which were all up;
the fixed spins are
randomly chosen in the present calculation.}
At each concentration, the effective stiffness exponents,
$\aU (c)$, $\aS (c)$ and $\aR (c)$,
are defined from the defect energies
$\WUL (c)$, $\WSL (c)$ and $\WRL (c)$, respectively,
as $W_L (c)\sim L^{a(c)}$.
Here suffixes $\rm U, S, R$  distinguish the BCs, and
$L$ indicates the linear size of the system.
Since each stiffness exponent is expected\cite{Ozeki90,UenoO91}
to behave as
$a>0$ in the ordered phase,
$a=0$ at the critical point
and $a<0$ in the disordered phase,
one can distinguish the phases according to the signs of
$\aU$, $\aS$ and $\aR$.

Calculations are carried out at several concentrations
for the square lattices of linear size
$L=6, 8, 10, 12, 14, 16$.
At each concentration, $W_L$ is estimated by averaging
$10000-70000$ randomly chosen samples of ion configurations.
The concentration dependence of the stiffness exponents is shown in Fig. 2.
Then, we estimate the critical concentrations as
$\ccF \approx 0.63 \pm 0.01$ and
$\ccA \approx 0.41 \pm 0.01$.
Since $\aR (c)$ is always positive for $0\le \cA \le 1$,
we have evidence of the existence of the SG phase
in $0.41 <\cA <0.63$.\footnote{
Although one should show some typical dynamical behaviors to confirm
the SG phase,
the above behaviors are sufficient in the static sense.}

As shown rigorously, $\aR (c)$ is symmetric with respect to $\cA =1/2$.
On the other hand, the critical concentrations for the FM and the AF phases
are slightly asymmetric.
This indicates
the existence of the SG phase at least in $0.59 <\cA <0.63$,
which is consistent with the behavior of $\aR$.
Accordingly, we obtain evidence of the SG phase
in a two-dimensional Ising system
from both the conventional analysis and a new criterion.
However,
we should carefully conclude this point, because
the concentration difference of widths of two phases
($0.04 \pm 0.02$) is too narrow, and
the system size is still small to confirm
such a sensitive property.

Note that both regions of the FM and AF phases in the numerical results
are larger than those in the $\pm J$ mapping.
In particular,
the AF phase remains up to $\ccA\approx 0.41$,
where $-J$ bonds are distributed less than $+J$ bonds.
Let us discuss the reason for this.
At the critical concentration, $\cA\approx 0.41$,
the concentration of the B-ion ($\approx 0.59$) is
close to the percolation threshold of the site process in 2d,
$0.592745$.\cite{Stauff85}
Since the $-J$ bonds are located only between B-ions,
the infinite percolated cluster of B-ions is
connected antiferromagnetically
in the remnant FM bonds for $\cA < 0.407255=(1-0.592745)$.
If the remnant bonds have no exchange interaction, {\it i.e.},
the site-diluted system, the AF phase exists for
$0\le \cA\le 0.407255$ in the ground state.
However, there are two differences
between the present system and the site-diluted system.
One is the existence of frustrations,
which usually inhibits the long-range order.
Another is the fact that
the clusters are not magnetically independent of
each other in the present system.
However, these two facts are not expected to change this
``percolated-cluster picture" for the following reasons.

In the $\pm J$ model, the FM order is destroyed
at $p=0.89$ much
faster than the percolation threshold of the bond process ---
$p=1/2$ on a 2d square lattice.
The degree of frustration in the present system
is weaker than that of the $\pm J$ model:
Since $\pF \approx 4\cA^2 (1-\cA )^2 $,
the maximum concentration of frustrated plaquettes is
$\pF =1/4$ at $\cA=1/2$,
while $\pF =1/2$ at $p=1/2$ in the $\pm J$ model.
Even at the critical concentration
$p=0.89$ or $p=0.11$ in the $\pm J$ model,
$\pF =0.317$
is greater than the maximum value $\pF=1/4$
in the present model.
Thus, the concentration of frustrated plaquettes is always
lower than that at the FM (AF) critical point of the $\pm J$ model,
at which the frustration just destroys the uniform (FM) order.

In spite of the interactions between clusters,
the effect through the boundaries of clusters is
not expected to be strong.
In the present site-random system, frustrations do not exist inside
the clusters but among them.
Typical connections of two adjacent FM and AF clusters
are drawn in Fig. 3.
If the boundary of the clusters is oblique to all the axes as in Fig. 3(a),
no frustrations appear between the clusters
and the magnetic orderings of two clusters do not compete.
If it is parallel to one of the axes as in Fig. 3(b),
all plaquettes between the clusters are frustrated, and
each cluster can flip without changing the energy.
If these two cases appear separately, the long-range order in an infinite
cluster is stable.
Although these two cases generally occur together,
the effect of inhibition would be smaller than that in the $\pm J$ model.
In fact, the FM and AF phases are wider than those obtained
by the $\pm J$ mapping.
Moreover, the AF ordering in the $+J$-bond-rich region
comes from the percolated cluster of B-ions.
The reason of the discrepancy between
the FM and AF critical concentrations is not clear at present, and
should be solved in future.

In two dimensions, the region of infinite percolation for the A-ion
has no overlap with that for the B-ion.
In three dimensions, the percolation threshold of the cubic lattice is
0.3117,\cite{Stauff85}
and infinite clusters of both A-ions and B-ions exist
for $0.3117 <\cA <0.6883$.
In this region, the FM order on the A-ions and the AF order on the B-ions
strongly compete, and consequently the SG phase would appear.
Therefore, even if the SG phase exists in two dimensions,
its origin may be qualitatively different from that of the SG phase
in three dimensions.

In summary, we have studied the site-random Ising spin glass model.
Although this model has no gauge symmetry in contrast to the $\pm J$ model,
we find a new rigorous symmetry for the SG correlation
and the free energy.
This symmetry provides restrictions on
the structure of the phase diagram, and
a criterion for the existence of the SG phase
from a different point of view to that of conventional arguments.
We calculate the defect energies with uniform, staggered and replica
boundary conditions in the ground state in $d=2$
using the numerical transfer matrix method,
and obtain evidence of the SG phase from both
the conventional scaling analysis and the new criterion.
We propose the percolated-cluster picture in order to explain
the transitions from the FM and the AF phases in the ground state,
which are expected to be the origin of the SG ordering in $d\ge 3$.
Thermodynamic properties and critical behaviors
of the present site-random model are
quite different from those of the bond-random models
which have been investigated as the models of SG materials.
Further studies in this direction are required.

Numerical calculations were carried out on FACOM VPP500 at the Institute of
Solid State Physics, University of Tokyo.
One of the authors (Y. N.) is grateful for financial support from the
Japan Society for the Promotion of Science for Japanese Junior Scientists.
%
%
%


\begin{figure}[h]
\caption{Schematic phase diagram of the site-random SG model.
The asymmetry of the FM and the AF phases is assumed.
The shaded area indicates the excess region out of the FM phase,
in which the SG phase exists.
\label{fig1}}
\end{figure}

\begin{figure}[h]
\caption{
Stiffness exponents $\aR$, $\aU$ and $\aS$
as a function of the A-ion concentration $\cA$.
These exponents are evaluated from the defect energies
of $L=6\sim 16$ with
the replica, uniform and staggered boundary conditions, respectively.
\label{fig2}}
\end{figure}

\begin{figure}[h]
\caption{
Schematic ion configurations at the boundary of the A-ion (closed circles)
and the B-ion (open circles) clusters.
Bold and dotted lines indicate $+J$ and $-J$ exchange interactions,
respectively.
Frustrated plaquettes are marked by crosses,
while unfrustrated ones are unmarked.
\label{fig3}}
\end{figure}

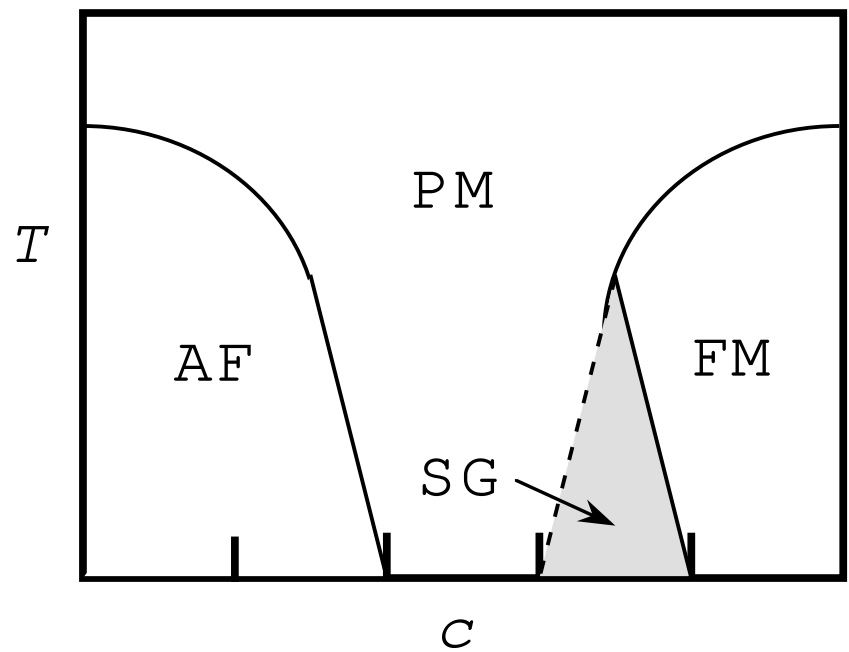
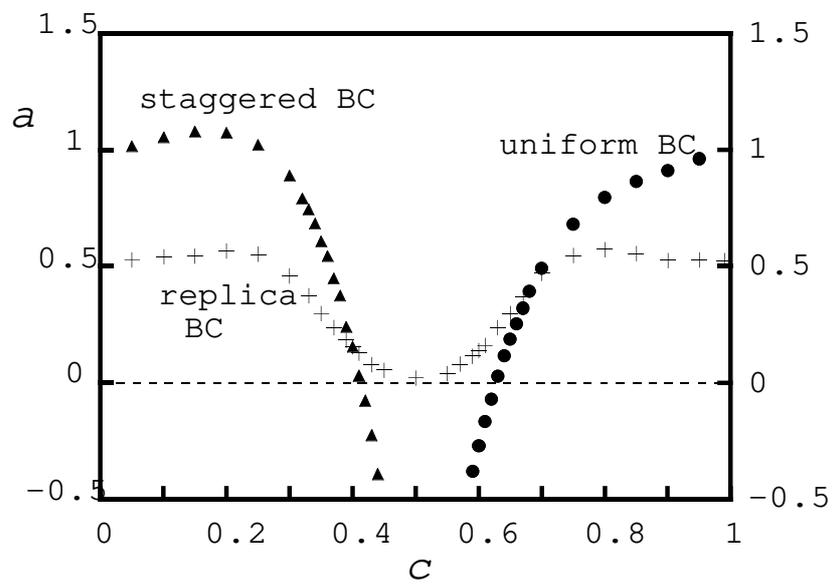
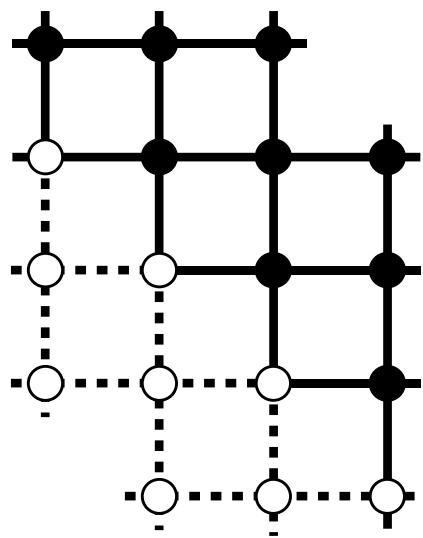
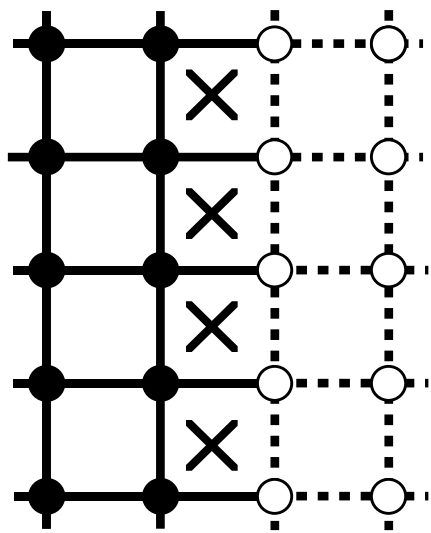
\end{document}